# Population Inversion Induced by Landau-Zener Transition in a Strongly Driven rf-SQUID

Guozhu Sun,[1,*] Xueda Wen,[2] Yiwen Wang,[2] Shanhua Cong,[1] Jian Chen,[1]
Lin Kang,[1] Weiwei Xu,[1] Yang Yu,[2,†] Siyuan Han,[1,3] and Peiheng Wu[1]

[1]*Research Institute of Superconductor Electronics and Department of Electronic Science and Engineering,
Nanjing University, Nanjing 210093, People's Republic of China*
[2]*National Laboratory of Solid State Microstructures and Department of Physics, Nanjing University, Nanjing 210093, China*
[3]*Department of Physics and Astronomy, University of Kansas, Lawrence, KS 66045, USA*

Microwave resonances between discrete macroscopically distinct quantum states with single photon and multiphoton absorption are observed in a strongly driven radio frequency superconducting quantum interference device flux qubit. The amplitude of the resonant peaks and dips are modulated by the power of the applied microwave irradiation and a population inversion is generated at low flux bias. These results, which can be addressed with Landau-Zener transition, are useful to develop an alternative means to initialize and manipulate the flux qubit, as well as to do a controllable population inversion used in a micromaser.

As controllable artificial atoms, superconducting qubits have received considerable attention because of providing a new paradigm of quantum solid state physics. So far, many fantastic macroscopic quantum coherent phenomena[1,2] have been demonstrated in superconducting qubits. In addition, recent experiments show that the interaction between superconducting qubits and microwave (MW) resonant cavity can produce single photon[2], which leads to a possible application of superconducting qubits as a micromaser. It is well known that population inversion, which maintains a majority of atoms in excited states rather than in ground state, has to be realized in order to ensure the amplification of the light and thus the laser process. Previous work[3,4] suggests that it is possible to generate population inversion in the superconducting quantum circuits subjected to MW radiation. In this letter, we report a further step in this direction: a controllable population inversion in a strongly driven radio frequency superconducting quantum interference device (rf-SQUID) by employing LandauCZener (LZ) transition.

LZ transition is a celebrated quantum mechanical phenomenon in the quantum world[5–7]. It has been found in various physical systems, such as atoms in accelerating optical lattices[8,9], superlattices[10], nanomagnets[11,12], quantum dots[13], and Josephson junctions[14–16]. Recently LZ transition is also found in the superconducting qubits[17–20], providing new insights into the fundamentals of quantum mechanics and holding promise for the superconducting qubits application. One may use LZ to enhance the quantum tunneling rate[21,22], prepare the quantum state[23], control the qubit gate operations[23,24] effectively, and do the controllable population inversion as demonstrated in this letter.

Our design of the superconducting flux qubit, being immune to charge noise and comparatively easy to be read out, is based on rf-SQUID[25]. An rf-SQUID consists of a superconducting loop with inductance $L$ interrupted by one Josephson junction with capacitance $C$ and critical current $I_c$. Its dynamics can be described in terms of the variable $\Phi$ and are identical to those of a particle of mass $C$ with kinetic energy $C\dot{\Phi}^2/2$ moving in the potential $U(\Phi)$. Here

$$U(\Phi) = U_0\left\{\frac{1}{2}\left[\frac{2\pi(\Phi - \Phi_f^q)}{\Phi_0}\right]^2 - \beta_L \cos(\frac{2\pi\Phi}{\Phi_0})\right\}, \quad (1)$$

where $\Phi_0$ is the flux quantum, $U_0 = \Phi_0^2/(4\pi^2 L)$ and $\beta_L = 2\pi L I_c/\Phi_0$. For $1 < \beta_L < 4.6$, when a magnetic flux $\Phi_f^q$ close to one half of a flux quantum is applied, the potential is a double well potential and the lowest states in each of the double wells serve as the qubits states. The states in different wells correspond to macroscopic current circulating around the loop with clockwise and counterclockwise directions, which can be distinguished by a direct current SQUID (dc-SQUID) magnetometer. Due to its large geometric size, the quantum phenomena found in the rf-SQUID are really macroscopic.

Our samples are fabricated with $Nb/AlO_x/Nb$ trilayer on an oxidized Si wafer using a standard photolithography process. An optical micrograph and schematic of one of the designs is shown in Fig. 1. The qubit is essentially an rf-SQUID in second order gradiometric design. Two tunnel junctions in parallel are used instead of one in order to adjust $I_c$ with an applied flux $\Phi_f^{CJJ}$, which consequently change the barrier height of the well *in situ*. We use an on chip flux bias line to control the tilt of the potential with an applied flux $\Phi_f^q$. Additional flux bias line is used to produce $\Phi_f^{dc}$ thus biasing the readout dc-SQUID at the maximum sensitivity region. The mutual inductance between the qubit and dc-SQUID is well designed so that we can distinguish the different flux generated by the circulating currents in the qubit. The sample is mounted on a chip carrier enclosed in a superconducting aluminum sample cell. The detailed description of the system is in Ref.26. The device is thermalized at $T = 20mK$.

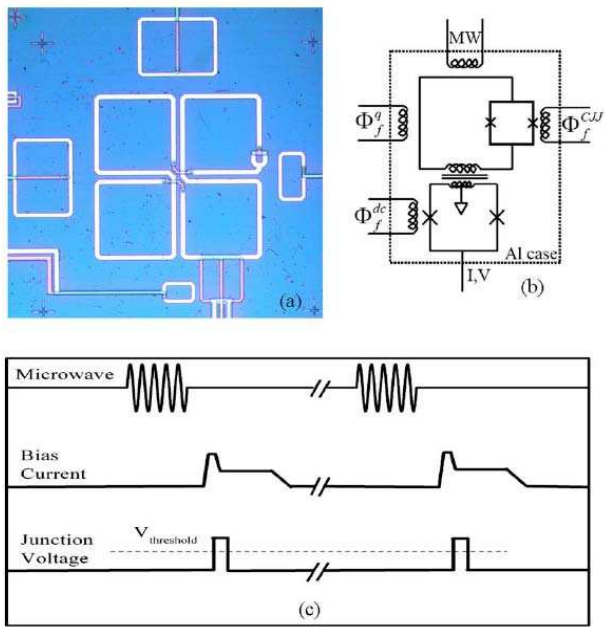

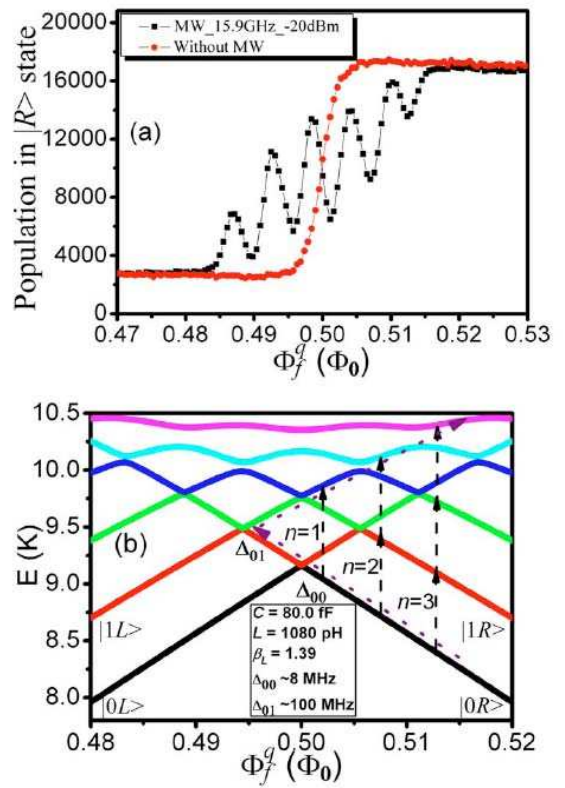

FIG. 1: (Color online) (a) Optical micrograph of the sample. (b) Schematic of the manipulation and measurement of the qubit. (c) A schematic of the time profile of the manipulation and measurement procedure.

FIG. 2: (Color online) (a) Example data show the dependence of the population in $|R\rangle$ state on the qubit flux bias without MW irradiation(dots) and with MW frequency f=15.9GHz and nominal MW power P=-20dBm(squares), respectively. (b) Energy-level diagram calculated using the parameters in our experiments. The dashed arrows indicate the position of the resonant dips observed in the experiments as examples.

The time profile of the manipulation and measurement sequence is shown in Fig.2. After calibrating the sample parameters, we keep the flux bias $\Phi_f^{CJJ}$; hence the barrier of the well is at an optimal value. The qubit is biased at a flux $\Phi_f^q$ near $0.5\Phi_0$. A MW generates a sinusoidal electromagnetic flux in the qubit and produces transitions between quantum states. A readout current pulse is applied to dc-SQUID shortly after MW irradiation is shut down. By setting the amplitude and duration of bias current pulse properly, the dc-SQUID either switches to finite voltage or stays at zero voltage corresponding to the qubit states being in left well $|L\rangle$ or in right well $|R\rangle$. The switching probability which can be obtained by repeating the trials $2 \times 10^4$ times, thereby represents the population in $|R\rangle$ state. Changing $\Phi_f^q$ gradually and repeating the above measurement, we obtain the population in $|R\rangle$ state as a function of the flux bias at fixed MW frequency and power(see Fig. 2(a)). Then by tuning MW power, we get the dependence of the population in $|R\rangle$ state on $\Phi_f^q$ and MW power.

Shown in Fig. 2(a) are example data for MW frequency $f = 15.9 GHz$ and nominal MW power $P = -20 dBm$. Without MW irradiation, the qubit relaxes to the ground state. There is a steplike transition near $0.5\Phi_0$, indicating the change in ground state from $|L\rangle$ to $|R\rangle$ at $0.5\Phi_0$. When irradiated by continuous MW ($1\mu S$ width) matching the energy-level spacing, peaks (dips) appear in the curve (squares), corresponding to MW induced excitation from $|L\rangle$ to $|R\rangle$ (from $|R\rangle$ to $|L\rangle$). The positions of all peaks and dips agree with the energy-level diagram calculated using the parameters $L = 1080pH$, $C = 80pF$, and $\beta_L = 1.39$, which are determined from independent measurement (Fig. 2(b)). Since the power of the MW is relatively large, transitions due to two- and three-photon absorptions can also be observed. These multiphoton absorptions are reported before in persistent current qubits[17,18,27]. However, the dimension of our rf-SQUID is much larger than the persistent current qubit. In addition, the driving frequency is large in our experiment. Therefore, large MW powers are needed to observe multiphoton transition.

The population in $|R\rangle$ state versus qubit flux bias $\Phi_f^q$ and MW power at frequency $f = 15.9GHz$ is shown in Fig. 3(a). It is interesting that the population is not a monotonic function of the MW power. The simple photon induced transition picture of a two level system predicts that the population on the excited states will increase monotonically from 0 to the saturate 0.5 with MW power. Therefore, our results suggest more than two levels were involved into our rf-SQUID qubit. We use notations $|iR\rangle$ and $|iL\rangle$ to represent the $i$th states in right well and left well, respectively, and model the strongly driven qubit by the four-level Hamiltonian in the basis of diabatic states $\{|1R\rangle, |1L\rangle, |0R\rangle, |0L\rangle\}$

$$\hat{H}_0 = \hbar \begin{pmatrix} E_{1R}(t) & \Delta_{11} & 0 & \Delta_{01} \\ \Delta_{11} & E_{1L}(t) & \Delta_{01} & 0 \\ 0 & \Delta_{01} & E_{0R}(t) & \Delta_{00} \\ \Delta_{01} & 0 & \Delta_{00} & E_{0L}(t) \end{pmatrix}, \quad (2)$$

where $E_x(t) = E_{x0} + k_x \Phi_{rf} \sin \omega t$, $x \in (1R, 1L, 0R, 0L)$, $\Delta_{i,j}[i,j \in (0,1)]$ is the interwell tunneling splitting(see Fig.2(b)), $E_{x0}$ is the detuning proportional to dc flux bias, $\Phi_{rf}$ is the amplitude of the periodic driving signal (i.e., MW), $\omega$ is the MW frequency, and $k_x = dE_x/df$ is the diabatic energy-level slope of state $x$. The parameters are shown in Fig. 2(b). The quantum dynamics of the system, including the effects of various decaying rates, is described by the Bloch equation of the time evolution of the density operator[28]

$$\dot{\rho} = -i[\hat{H}_0, \rho] + \Gamma[\rho]. \quad (3)$$

The second term $\Gamma[\rho]$ describes the relaxation and dephasing processes due to the environment's dissipation.

Figure 3(b) shows the simulated population using Eqs. (2) and (3) with parameters discussed above. Both single photon and multiphoton absorption due to LZ transition are clearly observed. The population is also modulated by MW power with strong population inversion. The agreement with the experimental results is remarkable.

The MachCZehnder interference with a maximum (minimum) in the resonant peaks (dips) generated by LZ transition are reported before[17,18]. However, unlike the result in the previous work, the interference fringes in our experiments do not exhibit a clear Bessel function dependence on MW power. One reason of the difference is the short decoherence time of our qubit since our qubit couples strongly with the environment due to the larger size. A rough estimation[29,30] gives intrawell relaxation rate $\Gamma_1 \sim (1ns)^{-1}$, interwell relaxation rate between $|0R\rangle$ and $|0L\rangle$ $\Gamma_{inter} \sim (1\mu S)^{-1}$, and dephasing rate $\Gamma_2 \sim (0.5nS)^{-1}$, which are appropriate for our qubit. Another reason of the difference is that we used an order of magnitude higher MW frequency. The modulation period of the population is proportional to the ratio of MW amplitude and the frequency. For a larger frequency we need larger amplitude to get the same modulation. However, with increasing MW power, the nonlinear or other high power effects will be dominant. This will degrade the perfect Bessel function dependence on the MW amplitude.

In addition, for a large MW frequency, dominant LZ transition occurs at the $\Delta_{01}$ crossing[31] instead of the $\Delta_{00}$ crossing because $\Delta_{00} \ll \Delta_{01}$ and the rate of LZ transitions is proportional to $\Delta_{ij}^2$.[18] By choosing a large MW frequency, one can easily reach the second diamond region in Ref. 31 where the population inversion can be generated. In our experiments, strong population inversions are generated with the maxima of 90% and 75% for

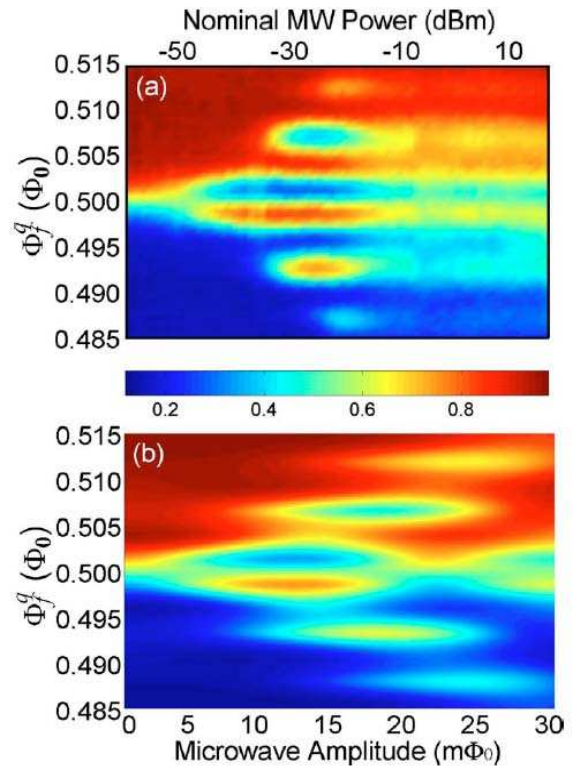

FIG. 3: (Color online) (a) 3D view of the dependence of the population in $|R\rangle$ state on the bias flux and MW power. X, y, and z axes represent MW power, the qubit flux bias $\Phi_f^q$, and the population in $|R\rangle$ state, respectively. MW frequency is 15.9 GHz. Strong population inversion due to LZ transition is clearly demonstrated. (b) The population in $|R\rangle$ state as a function of flux bias and MW power simulated using parameters in our experiments.

n=1 and n=2, respectively, which are due to the competition between transitions to the respective excited states $|1R\rangle$(or $|1L\rangle$) combined with fast intrawell relaxation to $|0L\rangle$ (or $|0R\rangle$). Since the MW power and frequency can be adjusted conveniently, one may realize a precise controllable population inversion in a superconducting qubit.

In summary, we have generated transitions between macroscopic quantum states in a strongly driven superconducting rf-SQUID flux qubit. The single photon and multiphoton absorption process have been observed. The population on the excited state is modulated with MW power. In addition, population inversion can also be realized, which is consistent with the LZ transition theory. Therefore, using high MW frequency and power, we may do the controllable population inversion in superconducting flux qubit. This enables one to envision a micromaser based on macroscopic quantum transitions.

This work is partially supported by NSFC (Grant Nos. 10704034 and 10534060), the State Key Program for Basic Research of China (Contract No. 2006CB921801), the Natural Science Foundation of Jangsu Province (Contract No. BK2007713).